# Network Parameters Impact on Dynamic Transmission Power Control in Vehicular Ad hoc Networks

KHAN Muhammad Imran

IRIT, University of Toulouse III, France

*Abstract*

*In vehicular ad hoc networks, the dynamic change in transmission power is very effective to increase the throughput of the wireless vehicular network and decrease the delay of the message communication between vehicular nodes on the highway. Whenever an event occurs on the highway, the reliability of the communication in the vehicular network becomes so vital so that event created messages should reach to all the moving network nodes. It becomes necessary that there should be no interference from outside of the network and all the neighbor nodes should lie in the transmission range of the reference vehicular node. Transmission range is directly proportional to the transmission power the moving node. If the transmission power will be high, the interference increases that can cause higher delay in message reception at receiver end, hence the performance of the network decreased. In this paper, it is analyzed that how transmission power can be controlled by considering other different parameter of the network such as; density, distance between moving nodes, different types of messages dissemination with their priority, selection of an antenna also affects on the transmission power. The dynamic control of transmission power in VANET serves also for the optimization of the resources where it needs, can be decreased and increased depending on the circumstances of the network. Different applications and events of different types also cause changes in transmission power to enhance the reachability. The analysis in this paper is comprised of density, distance with single hop and multi hop message broadcasting based dynamic transmission power control as well as antenna selection and applications based. Some summarized tables are produced according to the respective parameters of the vehicular network. At the end some valuable observations are made and discussed in detail. This paper concludes with a grand summary of all the protocols discussed in it.*

## Keywords

*VANET, Transmission Range, Transmission Power, DSRC, Density, Single Hop, Multi Hop, Applications, Omni and Directional Antennas, Reliability, Scalability, Delay, Throughput, Connectivity*

## 1. Introduction

Vehicular ad hoc networks are particular type of mobile ad hoc networks but with different dynamics of topology; such as speed of vehicle, geographic dynamics, dimensions of its vehicular node etc. Vehicular ad hoc networks constitutes of moving vehicular node on the road, as like on the highways, urban areas or rural areas etc. The communication between moving vehicular nodes plays an important role in intelligent transportation system. Communication is possible between vehicles within each other's transmission range, and with fixed gateways along the road for vehicular to infrastructure communication. The ability of vehicles to communicate directly with each other via wireless links and form ad hoc networks that produce the exciting applications. In particular, these networks have important applications in Intelligent Transportation Systems (ITS). Many of these applications require re- liable and efficient dissemination of traffic and road information via ad hoc network technology. This is, however, a difficult task due to the highly dynamic nature of these networks which results in their frequent fragmentation into disconnected clusters that merge and disintegrate

                                                                  1

dynamically. Efficient utilization of bandwidth of the radio resources deployed on the vehicular nodes could perform an important role to overcome the issues of reliable dissemination of information in the network.

In vehicular communication, message dissemination occurs between vehicle-to- vehicle (V2V) and also vehicle-to-infrastructure (V2I). ITS aims to provide drivers with safer, more efficient and more comfortable journeys. It could provide drivers with timely traffic congestion and road conditions information so that drivers can avoid congested or dangerous areas that could cause the hazards or delays etc. Vehicle-to-vehicle communication is also referred as Inter Vehicle Communication (IVC) that needs no infrastructure for communication between vehicles and each vehicular node is equipped with a wireless radio by which it can send and receive its own messages and forward messages for the other vehicles. Message dissemination in V2V is broadcasted in single-hop and multi-hop fashions. Broadcasting tech- nique is classified into four categories [1]; 1) *simple flooding,* 2) *probability based,* 3) *area based* and 4) *neighbor knowledge based*. The equipment used for vehicular com- munication is called on board unit (OBU) that consists of different components as such, wireless radio link, processing unit for the messages etc. Vehicular communi- cation helps to reduce the number of accidents and possible deaths by propagating messages prior to such accidents. Dedicated Short Range Communication (DSRC) has been regarded as the most promising technology applied by worldwide develop- ment for V2V communication [2] [3]. In 5.9 GHz band US Federal Communication Commission (FCC) has allocated 75 MHz spectrum band for DSRC. As most emer- gency messages are life critical, and should be delivered to other vehicles as fast and reliable as possible [4], the traditional broadcasting scheme without ACK mecha- nism is not suitable for emergency message delivery in IVC. Second, without an effective broadcast control in the network layer, multiple redundant messages may be exchanged among nodes, which could cause broadcast storm problem [5] and significantly degrade the network resource utilization.

Vehicles on the road encounter different traffic conditions, such as during traffic jams, accidents, traffic lights, peak rush hours, late night etc., results in dynamic changes because their different mobility behavior. In rural highways and during late night hours, vehicles move with high speed due to low density on the road and hence sparse ad hoc network is created where connectivity between vehicular node is a challenging task. The communication link between V2V communication remains active only for a short period of time. To make the communication reliable, the connectivity should last for long time between vehicular nodes. By increasing the transmission range for sparse ad hoc networks it could be achieved as the farthest node could access the channel to communication with other vehicles. It signify that tuning the transmission power in VANET is so much important to regulate some issue regarding the connectivity and timely message propagation from vehicle to vehicle. Some of work to adjust the transmission power has been done considering different goals to achieve is explained such as; Impact of transmission power on the performance of UDP packets transmission is presented in [6], Dynamic adaptation of transmission power with contention window size to transfer the packets with their priority selection has been proposed in [7]. To enhance the connectivity and the duration of the path lifetime between source vehicular node to destination is proposed by S.Y.Wang in [8]. Similarly in [9][10][11], authors proposed the dynamic adaptation rate control of transmission power in VANET for delay con- strained applications and to access the channel to send timely the information to the destination. Neighboring vehicular density largely affect the Quality of Service (QoS) of the network if we consider the fixed transmission power. By adapting the dynamic transmission power according to the needs of the vehicular network that has a dynamic topology as well, so many problems described could be overcome.

In vehicular ad hoc networks, the dynamic change in transmission power is very effective to increase the throughput of the network and decrease the delay of the communication. Whenever an event occurs, the reliability of the communication from vehicular node to other vehicular nodes becomes so vital so that event messages should reach to these nodes. More importantly, the connectivity between the moving nodes comes first to achieve the reliability in the network. There is a direct relation between the connectivity and the transmission power, less transmission power means that connectivity between the moving nodes is weak and vice versa. This paper analyzes that how transmission power can be controlled by considering different parameter of the network such as; density, distance between moving nodes, information message priority etc. Optimization of the network could also be met by the dynamic control of transmission power in VANET; where it needs, transmission power can be decreased and increased depending on the circumstances of the network. Transmission range theoretically in DSRC standard is 1000 meter and the data rate can change from 6 Mbps to 27 Mbps. Transmission power versus transmission range is also calibrated in [12]. In the summarized tables some symbols are used to describe the behavior of the parameters of the network. Plus sign '++' shows the increasing and minus sign '−−' show the decreasing or degrading behavior, '±' describes the dynamic changing (increasing and decreasing) behaviors, the equal sign '==' indicates that there is no change; neither increasing nor decreasing.

This paper is organized as section 2 describes different parameters impact on dynamic transmission power control, then section 3 gives details on the density and distance based transmission power control. Section 4 describes the message broadcasting based and section 5 details the antenna based analysis of transmission power control. Section 6 describes the application and event based dynamic control of transmission power of vehicular node on the highway. Section 7 gives the observations and discussion in detail and conclusion with perspective is given at end in section 8.

## 2. Vehicular Network Parameters Affects on DTxPC

Some of the parameters such as, density, distance, broadcasting type, antenna type etc. are discussed here and their impact on dynamic transmission power control. In other words, we can say that how the change in transmission power could be affected in the vehicular network considering different situations on the highway either in broadcasting, event occurring situation or any network condition changes etc. A brief discussion is carried on in below paragraphs.

Density is one of the parameter that affect the transmission of the packets. Due to high number of vehicular nodes, the transmission medium becomes congested and probability of higher number of collisions increases that could decrease the performance of the vehicular network. So many papers [6] [7] [8] [13] [14] etc. have been published that change the transmission power of the network according to the density of the network. The dynamic change in transmission power is carried out by an algorithm throughout the network as the number of vehicular nodes in the network increases and decreases time by time on the course of the highway. Density of the network can be classified into two groups; one is simply by counting the number of vehicular nodes and the other is number of clusters of vehicular nodes that constitutes the network. Dynamic transmission power control (DTxPC) could be affected by the density of the network in both ways; by considering simply number of vehicular nodes and also by clusters of vehicular nodes connected to the network. Distance is also an important factor that affects the transmission range of the vehicular ad hoc networks. If the nodes as far away from each other, the transmission link between the nodes will be weak due to reception of lower power transmission signals. The closer the nodes will be, the stronger the signal strength will be between them. Distance also affects the density of the network, the more the distance between the nodes, less will be the density and vice versa. Some authors worked [13] [11] etc. on how the performance of the network could be optimized by considering the distance between the nodes. Transmission power control can be changed dynamically considering the distance could be called distance based dynamic transmission power control.

Dynamically change in the transmission power of a vehicular node in a vehicular network can enhance the reliability of any application driven for the safety purpose or simply for the communication in the network. It can be happened for V2V or V2I network communication. Whenever some event happen in the network, the vehicular node could change their transmission power according to the impact of the event to communicate its information to maximum or desired length of the highway or to maximum or desired the number of moving vehicles. So dynamic transmission power control (DTxPC) could be event-driven and also an application based. In communication system, different messages have different priorities e.g. voice and video messages do not have the same priority. VANET communication messages could be of different type as like; emergency message, accident messages, warning message, etc. so these all have different priorities. The transmission power control could not be same all the time to communicate those VANET messages. During the course of communication between vehicular nodes, dynamic change in transmission power according to the messages priority [14] [15] etc. become vital for the efficient utilization of resources.

Different type of antennas are used in VANET for the wireless access between moving nodes in IVC and also in V2I network. Different antennas are deployed considering different applications in VANET. The most important need of vehicular network is to transmit the safety message to save the lives of the passengers. The authors used different type of antennas [10] [13] [9] etc. in vehicular ad hoc network communication. Omni-directional antennas are used for the coverage of 360 degree area and directional antennas are used to cover the particular direction on the highway. Dynamic transmission power control can also classified on the basis of an antenna selection. These antennas are used for broadcast wire-less communication between V2V and V2I. The message transmission in vehicular network could be single hop broadcast and also multi hop broadcast. Single hop broadcast communication occurs in the neighboring vehicular nodes which are in the transmission range of a particular reference vehicular node. Multi hop broad-cast communication travel from one transmission zone to several other transmission zone until the destination is reached. According to the broadcast need and also de-pending on other parameters such as distance, transmission power of the reference vehicular node is changed dynamically during the course of communication. If the nodes are closer, for single hop broadcast communication less power is needed for the reliable transmission and vice versa. In multi hop broadcast communication, to reach the destination in a minimum time, transmission power of a vehicular node could be increased so that maximum distance could be covered in less time. For the rapid and reliable communication, dynamic transmission power control (DTxPC) of a vehicular node plays a crucial role [16] [17] [18] etc. in a single and multi hop broadcast communication.

In the below sections, the detail description is done explaining with the algorithms and protocols designed accordingly, and also their impact on delay and throughput constraints and other essential parameters of the vehicular network.

## 3. Density and Distance based DTxPC

Vehicular equipped with On Board Units (OBU) constitutes the wireless vehicular ad hoc networks on the highway. These moving nodes communicate with each other through OBUs that contains also antenna for message propagation. Density of the vehicular network depends on the number of the moving nodes. If moving nodes are greater in number then density will be higher and vice versa. The distance between these moving nodes has an inverse relationship with the density in a confined area of the highway. If the density is high, the distance among the nodes will be low and vice versa. The transmission power plays an important role for the best performance of the vehicular network. Generally the fixed transmit power and the QoS related parameters for prioritized messages do not enhance the performance for the dynamically changing topology of VANETs. In order to achieve better performance utilizing the local density information, dynamically changing transmission power with dynamic adaptation of contention window size in EDCA is needed in V2V communication. Vehicles estimate the node density by gathering the neighbors information within the current transmission range. The moving nodes make the connection among themselves for better and reliable communication and get connected in the transmission range of each other. As density of the network increase and decrease on the course of the highway, to get reliable communication and lesser delay, dynamic transmission power is necessary for the vehicular nodes based on the local traffic density information rather than high fixed transmission power. The table ( 1 ) summarizes some protocols and algorithms proposed in the literature and these are also discussed with their impact on dynamic transmission power control and range estimation. Moreover, delay and throughput and other parameters are also discussed accordingly.

The power control algorithm proposed in [9] is capable of the managing the topology of a vehicular ad hoc network by adjusting transmission power dynamically. It is based only on local information and no exchange of power-related signaling among nodes is required. It is scalable for a very low density up to a very high user density and this is achieved by controlling the transmission power, so that the number of neighbors of each node is always within a minimum and maximum threshold. This power control algorithm proposes a mechanism that adjusts the transmission power adaptively based on number of neighbors. First each vehicle starts with initial transmission power $P_{ini} = -35dBm$. It incrementally increases the transmission power as long as the number of neighbors is within a minimum threshold, or it reaches maximum transmission power value i.e. $P_{max} = -24dBm$. The transmission power is decreased when the number of neighbors greater than maximum threshold. Otherwise transmission power remains the same if the number of neighbors is within minimum and maximum threshold.

The transmission range in protocol proposed by [16] remains fixed, it does not change dynamically but it incorporates the distance parameter for multi hop trans- mission of messages. Position based multi hop broadcast protocol (PMBP) is developed for emergency message dissemination in IVC. End-to-end delay for an emergency message reserving channel access by exchange of BRTS and BCTS packets from source to destination could be given by the following equation; where $T_f$ is the time taken at the first hop which normally is longer than the following hops and $T$ denotes one-hop delay at intermediate hops and $T_l$ represents the time spent at last hop.

$$T_v = T_f - (s - 1) \cdot T - T_l, s \geq 1 \quad (1)$$

Where $s$ is the ratio of the expected total number of vehicles on the highway to the expected average number of vehicles within the transmission $R$. When $s = 0$, all vehicles are within the transmission range of the source node, and the end-to-end delay can be obtained from $T_f$.

| Protocol | Density | Distance | Tx Power | Delay | Cost | Impact on Network |
|---|---|---|---|---|---|---|
| PCAHCA[9] | ++ | - - | - - | ++ | Scalability | Connectivity |
| PMBP[16] | ++ | - - | = | - - | Scalability | Reliability |
| DAJTPCW[7] | ++ | - - | - - | - - | Scalability | Reliability |
| ETxRPL[8] | ++ | - - | - - | ++ | Throughput | Connectivity |
| EAEP[13] | ++ | - - | - - | ++ | | Scalability |
| ROR[19] | ++ | - - | = | ++ | Overhead | Reliability |
| TxPIoUDP[6] | ++ | - - | ++ | ++ | Interference | Connectivity |
| DawrTxRC[11] | ++ | - - | - - | - - | Throughput | Reliability |
| StopN'Go[14] | - - | ++ | ++ | ++ | Delay | Connectivity |
| DB-DIPC[10] | - - | ++ | ++ | - - | Resources | Connectivity |
| RCMChP[20] | - - | ++ | ++ | - - | | Reliability |
| CLBP[17] | - - | ++ | = | - - | Scalability | Reliability |

Table 1: Density and Distance based DTxPC

Rawat et al. in [7] proposed an algorithm for joint adaptation of transmission power and contention window to improve the performance of vehicular ad hoc net- works by a cross-layer approach. To make the connection reliable for long time, the proposed algorithm adapts the transmission power dynamically rather than high and fixed transmission power; based on the estimated local traffic density informa- tion. The algorithm that they have developed use the transmission range that is calculated by the traffic flow theory and the prioritization of different types of messages according to their urgency and delay requirements. The transmission range $TR$ is determined by ;

$$TR = \min\{L \times (1 - K), \sqrt{L \times \ln(L)/K}\} - \alpha \times L \quad (2)$$

Where $L$ represents the length of road segment, $K$ represents the estimated vehicle density and $\alpha$ is a traffic constant from traffic flow theory. Vehicles estimate the node density by gathering the neighbors information within the current transmission range. The new transmission range is then derived by using the traffic flow model as explained in Eq (2). Here transmission power is calculated by mapping on the basis of transmission range [12].

Path population on the highway depict the density of the vehicular network. In a protocol proposed by Wang [8], the path population of the highway is described in both direction of the highway separately; the first population can be represented as $SameDir$ consists of all the paths whose source and destination vehicles move in the same direction and the second path population called $DiffDir$, consists of all the paths whose source and destination vehicles move in the different directions on the highway. It uses vehicular mobility traces to investigate how wireless transmission range can effect the path lifetime in an IVC network in $SameDir$ density population and also in $DiffDir$ density population.

Edge-Aware Epidemic Protocol (EAEP) proposed in [13] is for highly dynamic and intermittently connected VANET. The density of the network is considered as the cluster environment in this protocol. Vehicular traffic models can be classified into macroscopic and microscopic models. In macroscopic model traffic is treated as an incompressible fluid, characterized by average density $\rho$ (cars/km), average velocity $V$ (km/h) and average traffic flow $Q = \rho V$. In microscopic model each car is treated individually and its motion in time and space is described using the so-called car-following models which incorporates the behavior of drivers in traffic through simple parameterized distance and velocity dependent interactions between adjacent cars. It used different range of transmission power for dissemination of messages by using RTS/CTS mechanism.

Xiaomin Ma et al. proposes distributive robust scheme in [19] for DSRC one-hop safety critical services. The proposed scheme enhances broadcast reliability using dynamic receiver-oriented-repetitions (ROR) and mini-slot within DIFS in IEEE 802.11 for one hop emergency warning messages dissemination. When a vehicle sends out a first cycle of emergency packet, one or more nodes receive these packets and responsible to repeat the broadcast for their one hop neighbors. Exchange of beacon messages in the network results the update of node's location, mobility information of one hop neighboring nodes, moving direction, speed etc. Receiver node distinguishes the copies of broadcast packet from the newly generated packets through a 12-bit sequence number of the received packet in the MAC header of IEEE 802.11. The emergency message is reached to the last node in the transmission range $R$ of the sender vehicle. To estimate the time to reach the emergency message, an Assessing Delay *(AD)* is introduced as in the below equation (3); where, $T_{max}$ is maximum *AD* time duration allowed, and normally it is less than the message life time. $R$ is the communication range of the sender, and $d$ is the distance of the current node to the sender.

$$t_{AD} = T_{max}\left(1 - \frac{d}{R}\right) \qquad (3)$$

Vehicular nodes in this network model are displaced according to Poisson point process with density $\beta$ (nodes per meter) in a fixed transmission range $R$, the total number of nodes in the transmission range of reference vehicular node are $N_{tr} = 2\beta R$.

Transmission power affects the UDP performance in vehicular ad hoc networks is studied by Khorashadi et al. [6]. It is a simulation based work where they observed the change in throughput as by changing the transmission power in a given traffic density. So at certain point, the throughput remains flatten instead of increase in transmission power. It is because of interference at high frequencies. They studied the effect of dynamic transmission power in various traffic density and road scenarios on the UDP throughput. Experimentally they have established the fact the traffic density is only important at lower transmission range to provide the required connectivity. At low transmission power with low traffic density, the connectivity could not be stable. They found that the throughput has no clear correlation with vehicle traffic density than the transmission power. Thus increasing the transmission power reduces number of hops resulting in improved throughput.

The authors showed that dynamically tuning transmission power based on vehicle positions could be used to maximize throughput and decrease the number of hops.

Jialiang Li et al. in [11] investigates the impact of transmission range on the end-to-end delay in 802.11p-based vehicular ad hoc networks. It develops a concise expression of the transmission delay in saturated networks and obtains the service rate of non-saturated networks. The network will be saturated if the arrival rate of packets $\lambda$ is greater than service rate of packets $\mu$ and if $\lambda < \mu$ then network will be un-saturated. Density of the vehicles is considered to be homogeneous, so the number of vehicles in front is the same as the number of rear vehicles. *K* is the number of front vehicles within the communication range. To retransmit the message in a minimum time, the farthest vehicle is chosen e.g. $K^{th}$ vehicle from the sender. So $\lceil L/K \rceil$ gives the path length i.e. number of hops, where *L* is the source-destination distance that represents the number of intermediate vehicles. The per hop delay is the interval that a packet stays in each hop, including processing delay $D_{proc}$, queuing delay $D_{queue}$ and transmission delay $D_{trans}$, so the end-to-end delay can be expressed as

$$Delay = \lceil L/K \rceil \cdot (D_{proc} + D_{queue} + D_{trans}) \quad [11]$$

Rex Chen et al. in [14] uses the Stop-and-Go traffic waves with different traffic densities to measure the packet reception rate in different transmission ranges. If the transmission range is high then the interference factor degrades the performance of the network and the reliability of the network decreases. Stop-and-Go movement is a phenomenon that arises from a combination of shockwave and rarefaction waves, occur especially during peak hours or when some incident occurs on the highway. Dynamic transmission range has been adapted in [14] based on the traffic stability measures that achieves high reliability by considering network coverage and packet reception rates. Coefficient of Variance (*CV*) parameter of spacing between vehicles is determined by using the local information of each vehicles. Coefficient of variance has a inverse proportion to the vehicular density on the road. Transmission Range Adjustment is done by considering the traffic stability i.e. when *CV* has a fixed value. The increase in transmission range is relative to *CV* to ensure a desirable coverage value for all nodes in the road network for specific traffic pattern.

$$TR_{adj}(n) = (1 - n \times CV) \cdot TR_{avg-sp} \quad [14] \quad (4)$$

Where *n* is the order of magnitude for increasing the coefficient of variance *CV* and $TR_{avg-sp}$ is the average vehicle spacing over the entire traffic stream. When traffic becomes uniform i.e. $CV = 0$, $TR_{adj}$ is equal to $TR_{avg-sp}$.

Each vehicular node in Delay-Bounded Dynamic Interactive Power Control (DB-DIPC) algorithm proposed by C.Chigan et al. [10] adjusts its own transmission power in distributed way by exchanging the periodic Probe/data messages. In a message exchange duration defined as $T_{cycle}$, each vehicle sends only one *probe* or data message. After each $T_{cycle}$, vehicle evaluate its transmission power whether it is appropriate with other neighboring vehicles. It then decides how to adjust its own power and inform others with impropriate powers to adjust their transmission pow- ers by sending probe/data message. As the neighboring vehicles change the topology dynamically accordingly the node also enter into different stages of algorithm as probing stage *to* adjustment stage → stable stage → probing/adjustment stage•••. Latency performance (best, intermediate and worst) of DB-DIPC algorithm for vehicular nodes distributed in a Poisson point process; is derived. Therefore, $Delay = j \cdot T_{cycle}$, and the expectation of the delay $E(Delay) = E(j) \cdot T_{cycle}$. Expectation of delay can be expressed as $E(j) = E(j)_{best} + E(j)_{intermediate} + E(j)_{worst}$. Where *j* is number of steps to reach neighbor.

In Reliability of Cluster-based Multichannel MAC Protocol [20]; the author describes to find the cluster size and hence the communication range that maintains a high network stability and reliability, increases the life time of a path and at the same time decreases the time delay for an emergency message to reach its intended distance. All vehicular nodes in a cluster have a same communication range *R* and carrier sense range *CS*, i.e. they use the same transmitting power $P_t$ except the CH. Cluster head uses two levels of power, one level $P_t$ is the same as other members dedicated to communicate with its cluster members. The other power level to reach a distance *Dc* to communicate with the neighboring cluster heads. Communication range *R* of cluster head is also responsible to determine the cluster size and it determines the stability of the network as well. The communication range $R$ has to be selected on the basis of density of vehicles, status messages size and data rate such that all the cluster member could easily access the shared medium within CCI. If a vehicle has an emergency message, it will contend for the channel access using the minimum contention window specified for high priority class in IEEE 802.11p [21] to send this message for several times depending on the application. The emergency message will continue to propagate in the direction of interest for a maximum number of hops (M) depending on the application. Then the average time delay $T_{avg}$ for the emergency message to reach its intended distance of $M$ clusters is the sum of the time for the first cluster head to receive the message from its member, the time for the neighboring cluster heads to process and propagate the message and the time for the last cluster head to send the message to its members successfully.

$$T_{avg} = \left(\frac{1}{P_c} - \frac{M}{P_{cc}} + \frac{1}{P_s}\right) T_{data} + MT_p \quad (5)$$

Where, $T_{data} = \frac{L}{r_d}$, time needed to transmit the status message whose length is of $L$ bits with transmission rate of $r_d$ Mbps. $T_p$ is the time needed by the cluster head to process and analyze the emergency message before it propagates.

Cross Layer Broadcast Protocol proposed in [17] for emergency message dissemination in which relay node selection delays will be higher as increase of node density because of retransmissions caused by collisions. That node is more preferable for relaying an emergency message that has longer distance from source, better channel conditions and smaller relative velocity. CLBP delivers the emergency message as fast as possible. Number of hopes determines the how swiftly, the message could reach its destination. $\Delta d = \sqrt{(x - l_x)^2 + (y - l_y)^2}$ is a metric to determine the number of hops, the message will be forwarded with few number of hops with a larger $\delta d$. Small relative speed $\Delta v = |\vec{v} - \vec{v_r}|$ is usually desirable in high speed vehicle networks to guarantee the channel between two moving vehicles.

## 4. Message Broadcasting based DTxPC

Message broadcasting in vehicular ad hoc networks is particularly of two types, single hop broadcasting and multi hop broadcasting. In single hop broadcasting, the message is directly transmitted from the source vehicular node to its neighboring moving nodes. In multi hop broadcast, the message is transmitted to the destination vehicular node in a relay fashion by introducing intermediate moving node. If the transmission range is high, then, number of hops from source to destination node is less and vice versa. On highway, in a different situations of the vehicular networks, like density changes, events happening, messages priority etc, dynamic transmission power of the nodes optimizes the resources and also increase the performance of the network. Dynamic Transmission Power Control (DTxPC) affects the number of hops to transmit the event and safety messages in the network. In the below table (2) DTxPC is analyzed in different protocols and algorithms proposed in the VANET literature and explain briefly their mechanism.

The algorithm proposed in [9], where neighbors send and receive messages when they are in the transmission range of a reference's vehicle with certain transmission power depending on their locations. Single hop messaging is occurred when they interact wirelessly with each other. The transmission power that dynamically changes based on the number of neighbors but single hop broadcast communication between the nodes persists either DTxP decreases or increases.

In a fixed transmission power, in Position Based Multi-hop Broadcast Protocol (PMBP) [16]; adopts a cross layer approach considering MAC and Network layers. The highlights of this scheme includes: 1) by a cross layer approach, the current re- laying node selects the neighboring node with the farthest distance from the source node in the message propagation direction as the next relaying node, which ensures emergency messages can be delivered to remote nodes with the least time latency; 2) At each hop, the emergency message is only broadcasted once, therefore, redundant broadcast messages are greatly reduced; 3) by adopting revised BRTS/BCTS handshake, there is no hidden terminal problem in PMBP, and it ensures every node could correctly receive the emergency message, which makes the scheme more reliable; and 4) the emergency message has the highest priority to access the channel, and it guarantees the emergency message be broadcasted as soon as possible. The author modifies and adds some fields in RTS packet e.g. *position_ x*, *position_ y, em_info*. (*position_ x*, *position_ y*) is the current relaying node' position to broadcast the emergency message. *em_info* constitutes of source node address *init_ addr*, the emergency sequence number *em_seq* and the source node position (*init_x,init_y*). A vehicular node with an emergency message broadcasts a BRTS packet first, then starts the BRTS retransmission timer, if it does not receive a BCTS packet in that time, it will rebroadcast the BRTS packet until the rebroadcast time reaches $r_{max}$. The retransmission timer of BRTS packet can be given as below as $R$ represents the transmission range of a node and it is divided into a series of distance block denoted by *dis_slot* (actually its value should be the average length of vehicles);

$$t_{brts\_r} = t_{brts} + t_{sifs} + \left(\left\lfloor \frac{R}{dis\_slot} \right\rfloor\right) \cdot t_{slot} + t_{bcts} \qquad (6)$$

Rawat et al. in [7] propose an algorithm that used collision rate and the number of back off times as metrics to dynamically adapt the CW size. The main objective is to provide highest priority messages quickest channel access with higher trans- mission power so that safety messages can be broadcasted in larger region in single hop fashion. The values for CWmin, CWmax, and AIFS parameters are set based on the urgency level of the messages. In case of highest priority packets, the CWs and AIFS are set to smallest values and accordingly the transmission power is set to the maximum value. For other priority messages, the CWs and AIFs are set based on the priority levels and transmission power is set based on the node density.

In Edge-Aware Epidemic Protocol (EAEP) [13], the author introduce two access mechanisms for packet transmission; the basic access mechanism which uses an exponential back off procedure, and the RTS/CTS access mechanism. RTS/CTS is built on top of the basic access mechanism and uses a four-way handshake between source and destination in order to reduce bandwidth loss due to the hidden node problem. EAEP uses the probability to broadcast the information between vehicle- to-vehicle in omnidirectional and directional way on the highway. In omnidirectional propagation, vehicular node waits for a random time to make the decision upon receiving a new message, whether to rebroadcast it or not. The waiting time taken is exponentially biased towards vehicles which are further away from the source node and this waiting time is chosen between the interval [0, $T_{max}$], where;

$$T_{max} = \min \begin{cases} \frac{T_o}{U} exp\left(\frac{|x_{cur} - x_{snd}|}{L}\right) \\ T_{min} = \frac{T_o}{2U} \end{cases} \qquad (7)$$

U indicates the urgency parameter and L and To are also protocol parameters.
In ROR [19] when the emergency message and beacon messages are ready to transmit in the sender transmission range R, then a preemptive priority is given to emergency messages by setting up mini-slots, DIFS interval is devised into a number of mini slots as lm gives the length of mini-slot and $w_m$ represent number of mini-slots; $w_m = \lfloor DIFS/l_m \rfloor$

$l_m = 2\delta + l_{switch}$; where $\delta$ is the maximum propagation delay within the transmission range $R$, and $l_{switch}$ is the time duration that a transceiver switches between the receiving mode and the transmitting mode. The emergency message is first sent if its time is due after a short waiting time $t_m$, where $l_m \leq t_m \leq l_m w_m \leq DIFS$. The sender continues to send the emergency messages for a specific number of repe- titions ($N_c$) in a one-hop range until $T_{max}$ is expired. Otherwise, it switches to send the beacon message if the specific number of repetition are completed. The ROR scheme avoids the situation that results in possible continuous failures of message broadcast, and hence further enhances the broadcast reliability but it does not count the broadcast storm problem because of overhead of packets repetitions. One-hop One-cycle broadcasting of emergency message is during when the channel is sensed idle or just after the transmission of beacon message. The one-hop multi cycle emer- gency messages broadcast, multiple receiver-oriented repetitions are distributively performed through distance-based AD timer in each one-hop receivers.

Khorashadi et al. in [6] experimented that decrease in number of hops to trans- mit a message from a source to a destination is resulted by increasing the transmission range and hence the increase in throughput is observed.

| Protocol | Broadcast | TxPower | Throughput | IEEE MAC Scheme |
|---|---|---|---|---|
| PCAHCA[9] | Single Hop | -- | ++ | 802.11e |
| PMBP[16] | Multi Hop | = | -- | 802.11e |
| DAJTPCW[7] | Single Hop | -- | ++ | 802.11e |
| EAEP[13] | Multi Hop | -- | ++ | 802.11 |
| ROR[19] | Single Hop | = | ++ | 802.11 |
| TxPIoUDP[6] | Multi Hop | ++ | ++ | 802.11 |
| DawrTxRC[11] | Multi Hop | -- | -- | 802.11p |
| StopN'Go[14] | Single Hop | ++ | ++ | 802.11p |
| DB-DIPC[10] | Single Hop | ++ | ++ | 802.11 |
| RCMChP[20] | Multi Hop | ++ | ++ | 802.11p |
| EAR[18] | Multi Hop | -- | ++ | 802.11 |
| CLBP[17] | Multi Hop | = | ++ | 802.11e |

Table 2: Message Broadcasting based DTxPC

But it saturates the throughput when increase in transmission power occurred beyond a certain limit. If a node can not find a route to its destination, it will buffer the message and will check every *RETRANSMIT_DELAY* to see if the route has been found. Otherwise, the message will be dropped when *MESSAGE_TTL* will be over. In GPSR [22] the position-based routing, perimeter routing mode where it searches for the alternate routes that may not be geographical closer to the destination because of the dynamics of geographical area. The position of the next hop should be always closer to the destination node than that of the current hop. In every *BEACON_INTERVAL* seconds, every vehicular node broadcast its location to all its one hop neighbors. *INFO_TIMEOUT* time limit is used to update the location by periodic beacons otherwise, if any node could not transmit the beacon message, it will be considered out of the network. When a vehicular node wants to send any message to the des- tination, it will first broadcast its location information to its next hop neighbors even if its in beaconing interval. Further, the intermediate nodes also do the same; broadcast the location of original sender to their next hop neighbors. In the *BEACON_INTERVAL* second, the node also broadcast its position information; so while sending some message, it repeats the location update process that could increase broadcast storm problem in the network hence performance of the network could be slow.

The protocol proposed by Jialiang Li et al. in [11] where message dissemination is held in multi hop broadcast mode in IEEE 802.11p MAC scheme. The end-to-end delay is the summation of the per-hop delay along the path that means from source to destination. Per-hop delay is the interval in which packet stays in each hop, in- cluding processing delay, queuing delay and transmission delay. The per-hop delay is the interval that a packet stays in each hop, including processing delay $D_{proc}$, queuing delay $D_{queue}$ and transmission delay $D_{trans}$ [11].
Traffic Model of Stop-and-Go traffic waves in IVC [14] constitutes of one-lane highway unidirectional and vehicles are distributed in a non-uniform congested way. The message dissemination between the vehicles is in single hop broadcast fashion in the defined transmission range $R$.

Delay-Bounded Dynamic Interactive Power Control (DB-DIPC) algorithm pro- posed by C.Chigan et al. in [10] facilitates 1-hop connectivity via dynamic neighbor discovery. DB-DIPC interacts with the neighbors iteratively to adjust the network transmission power at run-time which consequently ensures that one hop neighbor connectivity to adapt the dynamics of vehicular network. DB-DIPC algorithm have basically three stages as probing stage, adjustment stage and stable stage and it repeats in a specific time period $T_{cycle}$ whenever a node discover a neighbor. So that neighbors could know the actual power levels of each other. If necessary, then neighboring nodes run the algorithm to adjust theirs transmission power accordingly. Transmission power changes by a function of power step exponentially as step function $step \leftarrow \lceil \log_2(power) \rceil$ is calculated in stable stage. In high transmissionpower it results high throughput.

Reliability of Cluster-based Multichannel MAC Protocol [20] uses the clusters on the highway to form the vehicular network. Cluster Head (CH) is responsible for the communication between other cluster heads, and it also broadcast the message to its cluster member. Any cluster member cannot by pass the cluster head to disseminate the information to any vehicular node which is the member of any other cluster. Firstly the author calculates the probability $P_s$ of the messages received by cluster members from their cluster head and then it calculates the probability $P_c$ of the status messages received by CH from its cluster members. Then $P_{cc}$ probability is calculated for the message propagation from one cluster head to another cluster head. The throughput of the vehicular network constituting of clusters increases. Transmission range of the cluster head determines the size of the cluster. If the size is greater, reception probability of the message will also be higher.

An Efficient Angular Routing protocol is proposed by S.Misra et al. in [18] which finds the minimum possible path length between a source and a destination involving minimum nodes to broadcast control and data packets. Information regarding the angular position of the nodes is exploited in selecting the most suitable node for transmission considering the proper network connectivity among nodes with minimum power consumption. Minimizing the number of total nodes involved during transmission lowers the overall power consumption rate and improvement of the throughput. EAR used angle-based routing to forward the data; instead of broadcasting the Route Request (RREQ) in the entire coverage of the node, the RREQ is sent only to those nodes who lie in a specific angle from source to destination node. EAR algorithm [18] works in a way that RREQ is generated, and if the existing routes are available in the cache, then it is forwarded. It investigates all the nodes' local information and find the nearest node to source that could retransmit the packets to the destination. Then it calculates the slope between that node and source node. By these two slopes' value, the angle is measured, then RREQ is forwarded to the destination between the area covered by that angle from the source.

Cross Layer Broadcast Protocol proposed in [17] for emergency message dissemination in a multi hop IVC network, aiming to improve the transmission reliability and minimize the message redundancy. For reliable transmissions of broadcast messages, broadcast request to send (BRTS) and broadcast clear to send (BCTS) frames are exchanged before sending of an emergency message. An appropriate relaying node is selected to forward the emergency message in the desired propagation direction. Before sending the emergency message, the nodes send the BRTS frame based on the CSMA/CA mechanism and wait for the response from any other node in its transmission range and starts the count for retransmission. When there is no BCTS response, the node contend for channel access to rebroadcast the BRTS immediately until a BCTS is successfully received. After successfully replying an ACK, the selected relay becomes the next relaying node and repeats the BRTS/BCTS handshake process.

Wenyang Guan et al. propose two adaptive message rate control algorithms for low priority safety messages in [15], in order to provide highly available channel for high priority emergency messages. First algorithm has two phases; In *Fast Start Phase* and *Congestion Avoidance Phase*, it uses threshold values to send the messages, if the messages count increases, congestion increases, periodic safety messages stops, and event safety messages dissemination continues. The second algorithm has only one phase called *Fast Recovery Phase* that is more aggressive in increasing message rate as compared to first algorithm it restarts from threshold $R_{thr}$ as compared to first algorithm who restarts from $R_{min}$. Vehicle wait for long time to increase the message rate when channel busy level approaches to the channel congestion threshold. Two algorithms provide a highly available low latency channel for event-driven safety applications (ESA) messages and for periodic safety application (PSA) messages.

## 5. Antenna based DTxPC

Dynamic Transmission Power Control depends on the antenna structure mounted on the vehicles moving on the highway. They are also the moving antennas as well. They serve for the radio range of the vehicular node for the wireless communication between V2V and V2I. Mostly two types of antenna are used, Omni directional antenna and Directional antenna. *Omni Directional* antenna transmit the signals in $360°$ direction while *Directional* antenna transmit the signals in a given direction. More than one antennas are also deployed on the vehicular node as in [23] for different purposes. Less channel congestion leads to high throughput and lesser delay for the message communication among the nodes which lie in the transmission range of observed vehicle. Below table (3) provides the summary of the protocols and algorithms proposed in the literature how dynamic transmission power control is affected by selecting the proper antenna model for vehicular communication.

The algorithm proposed in [9] uses the MAC multi frame consisting of three UTRA-TDD frame. A UTRA-TDD frame consists of 15 time slots. To access the slot, it is required to detect either it is *idle* or *engaged* or *collided*. If the received power at the receiver end is greater than the receiver's sensitive power range, then the physical layer assumes that channel is busy, otherwise, the channel is assumed to be idle. The MAC layer assumes that time slot is *idle* if it receive an idle indication from the PHY layer.

| Protocol | AntennaType | TxPower | RoadType | Corss Layer | Connectivity |
|---|---|---|---|---|---|
| PCAHCA[9] | Omni-directional | - - | Highway | MAC-PHY | - - |
| PMBP[16] | Directional | = | Highway | MAC-Network | ++ |
| DAJTPCW[7] | Omni-directional | - - | Highway | MAC-PHY | ++ |
| EAEP[13] | Both | - - | Highway | MAC-PHY | ++ |
| ROR[19] | Omni-directional | - - | Highway | MAC-PHY | ++ |
| DawrTxRC[11] | Omni-directional | - - | Highway | MAC-PHY | ++ |
| StopN'Go[14] | Directional | ++ | Highway |  | ++ |
| DB-DIPC[10] | Directional | ++ | Highway | MAC-PHY | ++ |
| RCMChP[20] | Omni-directional | ++ | Highway | MAC-PHY | ++ |

Table 3: Antenna based DTxPC

If MAC layers receives a data block, and busy indication from PHY layer, then MAC layer considers that time slot as *engaged*. If MAC layer does not decode the data block while busy indication of the time slot, then that time slot at MAC layer will be considered *collided*.

A directional antenna is used in PMBP [16] for a multi hop broadcast trans- mission to reach destination node. With a directional antenna a fix transmission power is set to transfer the emergency messages but with modified RTS/CTS packets based on IEEE 802.11e to support safety related applications. These packets are transmitted with a cross layer approach between MAC and Network layer.

Rawat et al. in [7] proposed algorithm for better performance of VANET net- work that uses omni directional antenna for dynamic transmission power over the length of highway with certain threshold value of density. It proposes a cross layer approach from physical to MAC layer. Transmission power adopts to accordingly the priority of the messages sent over the network.

EAEP [13] disseminate information in omnidirectional and directional way on the highway. In omnidirectional propagation, vehicular node waits for a random time to make the decision upon receiving a new message, whether to rebroadcast it or not. When the waiting time expires, vehicular nodes counts the number of messages received from front $N_f$ and backward $N_b$ directions. Based on the difference of these counts, the node makes a probabilistic decision whether to rebroadcast or not? The net effect of this scheme is that only nodes close to the edge of cluster keep message alive. In the case of directional message propagation, if a message is propagating forward/backward, it is only kept alive by nodes near the head/tail of the cluster.

In ROR [19] scheme, Omni directional antenna are used to transmit the messages on the channel in VANET. Every receiver end repeats the message transmission in a fixed transmission range $R$. The performance of the proposed scheme is measured on the highway vehicular networks by taking into account the impact of beacon message broadcast and fading channel conditions. Multiple repetitions of messages provide the redundancy of packet transmission in adverse DSRC environment which can enhance packet reception ratios but if it lasts for long time it may lead to continuous transmission failures because of collisions in the channel transmission.

Omni directional antenna used for delay-aware transmission range control mechanism by Jialiang Li et al. in [11] so the transmission range coverage is larger than the road width on the highway. The successful transmission from source to destination node is the time period to transmit a data packet. In saturated network the number of contenders are high to access the channel as compared to the un- saturated network. Therefore, collision probability also changes and hence a bit gain of throughput can occur in un-saturated network.

In Stop-and-Go traffic wave by Rex Chen et al. in [14] defined the coverage of the network. Measuring of communication coverage of the vehicular network with $n$ vehicles on the road with their particular locations, then the coverage of each vehicle $i$ is defined in terms of the Euclidean distance to the nearest upstream and downstream vehicles in the traffic stream. Transmission range of vehicle of $i$ can be denoted as $R_i$,

$$C_i = C_{i,upstream} + C_{i,downstream} \quad (8)$$

In Delay-Bounded Dynamic Interactive Power Control (DB-DIPC) algorithm pro- posed by C.Chigan et al. [10], the message transmission is directional over certain directional antennas. The architecture of DB-DIPC is based on a Relative Position Based MAC Nucleus (RPB-MACn) [23] which conceptually promises collision free wireless channel access, provided the transmission power is dynamically controlled within the directional 1-hop neighbors. So a directional antenna with a communication channel pair (Transmission and Reception) is dedicated for set of neighboring vehicles depending on their positions relative to the source vehicle. Since vehicles in different directions communicate using different antennas, the number of channel collisions is reduced.

Reliability of Cluster-based Multichannel MAC Protocol [20] proposes a net- work model that is built on one way multi lane highway where vehicles can only communicate with others which are moving in the same direction. Based on IEEE 802.11p, vehicles will alternate between control channel (CCH) and service channels (SCH) for every synchronization interval (*SI*). The time the vehicle spends on CCH called control channel interval (*CCI*) while time it spends on SCH called service channel interval (*SCI*), such that $SI = CCI + SCI$. Every element in any cluster is reachable by cluster heads. The size of the cluster head is governed by the communication range of the cluster head who has the role of dividing the channel between its members so that all elements have an access channel to send their status messages. CH also determine the sub-channel frequencies for all its members. If there are $N$ sub-channels, then probability to access the same sub-channel by two adjacent clusters is $1/N$. Clustering reliability is the probability $P = \sigma/CCI$ that a cluster member will transmit and receive the clustering information from its cluster head successfully. The carrier sense range $CS = \rho R$, where $1 \leq \rho \leq 2$, hence it will range from $R$ to $2R$.

## 6. Application, Event and Priority based DTxPC

Dedicated Short Range Communication (DSRC)[3] basically designed for two types of applications; one is safety applications and other is non-safety application. Vehicular ad hoc networks on the highway have both types of applications and moreover safety messages are divided into types, one is routine safety messages and other is event safety messages. Whenever any event occurs on the highway in VANET, it could either be happy or hazard moment and then nodes start to sending messages to other moving nodes to get well informed in the network. The messages that have high priority are processed first then based on priority table different messages are treated on their own turn. Fixed transmission power or range limit the scope of the message dissemination and hence the performance of the network is affected as compared to the dynamic changing in transmission power. Dynamic transmission power control could be on the basis of events happing on the course of the highway or any application initiation from any moving node. In the literature, different pro- tocols and algorithms are proposed that reveals the dynamic transmission power control enhances the performance of the network in terms of delay and throughput as well as connectivity and reliability of different applications. Summary of all these protocols and algorithms shown in table (4).

How the transmission power affects the User Datagram Protocol (UDP) perfor- mance in vehicular ad hoc networks is studied by Khorashadi et al. [6]. Decrease in number of hops to transmit a message from a source to a destination is resulted by increasing the transmission range and hence the increase in throughput is observed. The algorithm proposed in [9] describes the collision management when nodes are connected. Whenever an event occurs in the network, a mutual collision occurs when multiple reciprocally neighboring nodes transmit on the same time slot. It is treated as a simple collision when the users transmit on the same time slot and have a common reciprocal neighbor which is not involved in the collision. A reference node A assumes that a simple collision occurs in time slot *i* if it detects a busy channel during the time slot, but it can successfully capture one transmission.

PMBP adapts IEEE 802.11e to support safety related applications in IVC. Services are divided into five classes such that; Best effort, Video probe, Video, Voice and Safety. Different services have different priorities to access the channel based on the access categories (AC). The values of CW_MIN and CW_MAX are defined same as in IEEE 802.11b. After sensing the channel idle with the fix transmission power for AIFS interval, a node start its back off timer and then it computes the value of AIFS and contention window (CW). Higher the value of AC, the higher the priority of messages of that service. The value of the persistent factor set to 2.0 for all applications. The emergency messages have the highest priority to access the channel by adjusting AIFS, CW_MIN and CW_MAX.

Rawat et al. in [7] proposed an algorithm for joint adaptation of transmission power and contention window to improve the performance of vehicular ad hoc networks. It incorporates the contention based MAC protocol 802.11e enhanced distributed channel access (EDCA) to perform the timely propagation of high priority messages (e.g. emergency messages) in vehicle-to-vehicle communication. EDCA has the service differentiation to provide QoS for different type of messages such as voice traffic, video traffic, best effort traffic and background traffic and each of these have different AC ( access category) value. Higher the AC value, higher the priority. So the priority of the packets is adjusted by two parameter ; 1) transmission power level in physical layer and 2) MAC channel access parameters such as minimum contention window (CWmin), maximum contention window (CWmax), and arbitration inter frame space (AIFS). It always assign the maximum transmission range for the vehicle that carries high priority messages.

| Protocol | App or Event | Priority | TxPower | Throughput | BroadcastType |
|---|---|---|---|---|---|
| PCAHCA[9] | Event | High | ++ | ++ | Single Hop |
| PMBP[16] | App | High | = | - - | Multi Hop |
| DAJTPCW[7] | App | High | ++ | ++ | Single Hop |
| ETxRPL[8] | App | Medium | - - | - - | Unicast |
| ROR[19] | Event | High | - - | ++ | Single Hop |
| TxPIoUDP[6] | UDP App | Low | ++ | ++ | Multi Hop |
| StopN'Go[14] | App | Medium | ++ | ++ | Single Hop |
| ARCS[15] | App | High | = | ++ | Single Hop |

Table 4: Application Event and Priority based DTxPC

For unicast-based applications such as email, ftp and http etc. a long lifetime path is preferable rather than short lifetime; so by the mechanism proposed by [8] we can compute the expected lifetime of a path. The percentage distribution shows that as wireless transmission increases, more paths have longer lifetime and vice versa. If a path is established between source and destination in *N*'th second, then in each subsequent second, it checks whether the path connectivity still remains or broken. If the path is broken in *M*'th second, it tries to find the shortest backup path between source and destination vehicles, if there is no backup found, then path life of this repairable unicast path could be determined as *(M + 1) − N*. The quality of service metrics for real-time applications in vehicular ad hoc networks, this approach investigates the connection duration, route lifetime, and route repair frequency in VANET on the highway. Receiver Oriented Repetitions (ROR) [19] represents a mechanism for emergency messages in VANET where the broadcast message is reached all the nodes in a vehicular network area. To broadcast an emergency message in a form of packet when it occurs, vehicle send to control channel of DSRC a busy signaling to suppress the transmission of beacon messages and possible hidden terminals, so that emergency message could pass through channel on priority without any wait and collision. One-hop One-cycle broadcasting of emergency message is during when the channel is sensed idle or just after the transmission of beacon message. The One-hop multi cycle emergency messages broadcast, multiple receiver-oriented repetitions are distributively performed through distance-based AD timer in each one-hop receivers.

How the transmission power affects the UDP performance in vehicular ad hoc networks is studied by Khorashadi et al. [6]. UDP packets are used mostly in streaming audio and video applications that's why their work is application oriented in VANETs. But the reliability is the question about data dissemination in vehicular networks. The author experimentally investigated the effect of transmission power on UDP throughput. The throughput increases as the transmission power increases because the less number of hops are needed to reach the destination node in the network. Higher the transmission range, lesser the number of hops and vice versa.

Rex Chen et al. in [14] uses the Stop-and-Go traffic waves with different traffic densities to measure the packet reception rate in different transmission ranges. Packet reception rate is measured in the MAC level and is defined as the probability of receiving a packet sent within transmission distance. Vehicle safety applications on the highway with single-hop periodic broadcast communication which include pre-crash sensing and cooperative adaptive cruise control applications are the main achievement by adjusting the transmission range.
Adaptive Rate Control Scheme of DSRC based VANETs [15] deals with event safety application (ESA) and periodic safety application (PSA) by proposing two algorithms by Wenyang Guan et al.. The priority of the ESA is higher than PSA and the message rate is also increased with maximum utilization of the channel. The author address the QoS control issue from the aspect of safety message rate control, with objectives of providing high availability low latency (HALL) channel for event- driven safety applications (ESA) messages and try to maximize channel utilization for periodic safety application (PSA) messages. The proposed scheme utilize the local channel busy time as the indicator of network congestion and adaptively adjust safety message rate in a distributed way.

## 7. Observations and Discussion

VANET comes with unique characteristics such as, unlimited transmission power, predictable trajectory and plenty of potential applications. But it has also many challenges that include rapidly changing topology subject to frequent fragmentation and congestions, lack of connectivity, redundancy and stringent application requirement in real time robust message delivery. VANET could overcome the problem of transmission power when there is low number of neighboring vehicles i.e. vehicular traffic density, but with higher density, limited transmission power is not enough for the timely message propagation from source to destination vehicular nodes. High transmission power cause the interference problem that results in packet loss and low throughput with increased delay constraint. However; with high transmission power the connectivity can long last but the reliability of message dissemination decreases. The decrease in transmission power for high vehicle density or high pen- etration ratio and increase in transmission power for lower vehicle density or penetration ratio is proposed by several researchers in [7][9][10]. Transmission range in DSRC standard is 1000 meter. Road safety applications for VANETs require strict QoS and DSRC provide timely and reliable communication to make safety applications successful. QoS control for DSRC based vehicle safety applications is still an emerging research field and it can be achieved through mechanisms such as power control, message rate control and enhancing reliability etc. The challenge in implementing applications is in understanding of the complex dynamics of highly mobile single and multi-hop ad hoc networks which is a characteristics of VANET. For real time safety applications end-to-end delay of message delivery is very critical aspect that has been addressed and needs more attention to progress in VANET. The IEEE 802.11p or Wireless Access in Vehicular Environment (WAVE) has been adopted as a main technology for vehicular ad hoc networks (VANETs). Its Medium Access Control (MAC) protocol is based on the Distributed Coordination Function (DCF) of the IEEE 802.11 which has low performance and high collision rate especially when using a single channel. Therefore, many clustering based multi-channel MAC protocols have been proposed to limit channel contention, provide fair channel access within the cluster, increase the network capacity by the spatial reuse of network resources and control the network topology more effectively. Most of these protocols did not study the optimized cluster parameters such as average cluster size, communication range within the cluster and between cluster heads, and the life time of a path between highly mobile nodes. The table (5) shows the key features of each and every protocols and algorithm discussed and analyzed where ± sign indicates the dynamic transmission power and == indicates the fixed trans- mission power/range of vehicular nodes. In [11], the only focus is on the delay parameter in a multi hop way. But not considering the packet size limit, the above methodology do not provide the overhead of packets. Due to interference by the other vehicles that are out of transmission range could affect the reliability of the network. In saturated and non-saturated network conditions there is no clear way to find out the throughput of the network. Delay could be maximized by increasing the transmission range but on the cost of minimizing the throughput. But in [15] for adaptive message rate control maximize the channel utilization and hence in- crease the message reception rate i.e. throughput but on the cost of delay. Because delay can be increases as the number of vehicular nodes increase in the network and hence waiting time increases for unblocking the PSA message in MAC layer. It can risk to deliver the Emergency Safety Application (ESA) messages because they are queued when the congestion in channel occurs or message rate reaches its maximum threshold. As the threshold values are used for the density of vehicles and also for the collision in the channel, the protocol proposed in [7] does not seem to be scalable. Anyhow, the simulation results show the lower delays by prioritizing the different type of message and high throughput. With prioritization of messages in the system might not be able to satisfy the delay requirement of time sensitive high priority messages such as message related to an accident. In [8] The quality of ser- vice metrics for real-time applications in vehicular ad hoc networks, this approach investigates the connection duration, route lifetime, and route repair frequency in VANET on the highway. The simulations based approach for a rectangular highway uses the small transmission ranges from 100 to 150 m for a highway environment. This unicast approach does not clearly show the scalability of the network on the highway. If the path connection repairs itself several time during the whole life- time of the path, then delay and throughput constraints of the network could be affected adversely resulting in loss of information. The authors in [6] showed that dynamically tuning transmission power based on vehicle positions could be used to maximize throughput and decrease the number of hops. Hence, transmission range is inversely proportional to the number of hops. At higher transmission power causing interference that results in packets collisions, the important information could be lost that question the reliability of the network. But Caizzone *et al* in [9] shows that dynamically change in the transmission power, under low to high traffic densities; can control the number of vehicles under each vehicle's transmission range.

But proposed power control algorithm that is based only on local information and there is no exchange of power-related signaling among nodes. However, the design and dimensioning of efficient radio resource management policies should not only be based on system optimization aspects but also on traffic safety requirements. A potential drawback here is that the thresholds are static and do not reflect different vehicle traffic conditions and quality of road segments. Dynamic change in trans- mission power at run time without considering threshold values is proposed in [10]. In DB-DIPC the transmission power of the initiative node is verified iteratively and interactively by the neighbor nodes at run-time. The resulting transmission power for communications between immediate neighbors ensures that useful message signal is lower bounded and the associated interference effect is upper bounded; hence the transmission power converges into the optimal transmission zone enabling the small granularity of 1-hop neighbor connectivity promptly adapt VANET network dynamics at run time. DB-DIPC is a single hop that ensures the network connectivity but throughput of the network could be affected in a larger capacity of the network. As it considers its neighbor only to a 1-hop distance, the capacity and coverage problem of the network arises for the optimal resource utilization of the network.

Multiple repetitions in Receiver Oriented Reception [19] of messages provide the redundancy of packet transmission in adverse DSRC environment which can enhance packet reception ratios but if it lasts for long time it may lead to continuous transmission failures because of collisions in the channel transmission as the transmission power remains constant. As each and every node broadcast an emergency message to its one hop neighbors, the broadcast storm problem arises that affects network performance. Stop-and-Go Traffic Waves in IVC [14] considers only a specific traffic pattern while general traffic patterns on the highway have more than one lanes and because of their mobility dynamics, the communication pattern could be different than stop-and-go waves. Directional antenna with high transmission power and less density, reliability and coverage of the network increases as well as throughput. EAR algorithm [18] works in a way that RREQ is generated, and if the existing routes are available in the cache, then it is forwarded. RREQ contains IDs of source node and destination node. If it does not find any route , then it calculates by measuring the slope from source to destination. It investigates all the nodes' local information and find the nearest node to source that could retransmit the packets to the destination. When the node density is low, and the transmission range is also low, most probably there is less number of collisions. The message dissemination is also carried in the specific desired region rather than in 360◦ direction. There is an ambiguity if there will be more than one node found in the desired angle region then how EAR changes its behavior.

PMBP [16] and CLBP [17] are of multi hop broadcast protocols utilizing fixed transmission power to reach destination. They give the highest priority to the emergency safety messages. CLBP have less delay and less number of collisions as by introducing new BRTS/BCTS packets. In the position based multi hop broadcast protocol (PMBP), the farthest neighboring node waits the shortest time duration to reply the broadcast node and becomes the relaying node. The protocols proposed, choose the farthest neighboring node to forward the emergency message. However, due to the long communication distance, relative velocity, noise, etc., the farthest node usually has a bad channel condition and consequently achieves low transmission rate and suffers from high PER. High PER may cause MAC layer retransmissions and thus a long link delay for emergency messages. PMBP uses directional propagation with multi hop broadcasting that also increases the reliability. EAEP [13] is reliable, bandwidth efficient information dissemination based highly dynamic VANET protocol. It reduces control packet overhead by eliminating exchange of additional hello packets for message transfer between different clusters of vehicles and eases cluster maintenance. Dynamic changing transmission power in EAEP for cluster heads (CH) enhance the reliability and also control the packet overhead in the clusters. EAEP overcomes the simple flooding problem but it incurs high delay of data dissemination. In a very high density, the computation complexity can be increased. Selection of next forwarder during a random waiting time at instant of merging other cluster nodes can create the problem of packets overhead because of their identification by exchanging of beacon messages.

Table 5: Summarized DTxPC Table of Protocols

| Protocol | DTxPower | Snippet OR Resume |
|---|---|---|
| PCAHCA[9] | ± | High density, high throughput and also higher latency, Omni directional propagation with single hop broadcasting, high priority for event safety application on highway, low connectivity but non scalablity. |
| PMBP[16] | == | high density with lower delay and lower throughput, high priority to emergency messages, directional propagation of messages with multi hop broadcasting, increases reliability |
| DAJTPCW[7] | ± | Less density, high throughput, reliability increases, omni directional message dissemination with single hop broadcast, high priority for emergency safety messages |
| ETxRPL[8] | ± | Higher the transmission range, more the path repair time. enhances Reliability and Connectivity increase, not Scalable, its unicast dissemination. Less throughput, Higher delay. |
| EAEP[13] | ± | Omnidirectional and Directional propagation, probability based rebroadcast mechanism, scalable for macroscopic and microscopic, control packet overhead, high delay, enhances reliability and connectivity |
| ROR[19] | == | Omni-directional single hop, enhances reliability and connectivity with higher delay and higher throughput, overhead and packet repetition cause collision decreases performance, high priority to emergency messages |
| TxPIoUDP[6] | ± | Increase in transmission power, increase in throughput, high delay because of interference rise in low reliability, less density increase in connectivity, high transmission range lessens number of hops |
| DawrTxRC[11] | ± | Impact of transmission range on end-to-end delay in saturated and non-saturated network, lower latency, Omnidirectional on highway in multi hop broadcast mode, better connectivity, high collision in saturated network results low throughput |
| StopN'Go[14] | ± | Increase in density, decrease the spacing between vehicles, higher transmission range, higher throughput, directional communication on highway, high reliability and coverage |
| DB-DIPC[16] | ± | Decrease density, high transmission power, single hop with directional antennas, high throughput less delay, enhances connectivity and reliability but low coverage and lesser capacity |
| RCMChP[20] | ± | Lower delay and high throughput using cluster in the network, high reliability as well as connectivity between cluster head and cluster members, high priority for emergency messages, low transmission range results low collision to access channel |
| EAR[18] | ± | Lower transmission power, high throughput, lessens number of nodes to reach destination results less delay in multi hop, message broadcast in a specific measured area, low overhead packets, reduce collisions |
| CLBP[17] | == | Higher priority to emergency message, multi hop relay node selection, farthest node selection to rebroadcast, higher throughput, less density results less delay and less number of collisions |
| ARCS[15] | == | High packet reception rate, priority to emergency messages, threshold values of message rate, less collision with less latency, maximum channel utilization |

## 8. Conclusion & Perspective

This paper concludes that the dynamic change in transmission power is generally very effective to increase the throughput of the network and decrease the delay of the communication. Whenever an event occurs, the reliability of the communication from vehicular node to other vehicular nodes becomes so vital so that event messages should reach to these nodes. More importantly, the connectivity between the moving nodes comes in less and high transmission power is so vital for the reliable transfer of the messages among them. There is a direct relation between the connectivity and the transmission power depending on the density of the net- work, less transmission power means that connectivity between the moving nodes is weak and vice versa. In this paper, it is analyzed that how transmission power can be controlled by considering different parameter of the network such as; density, distance between moving nodes, message priority, antenna type and application or event initiation etc. Some summarized analysis tables are shown according to the respective parameters. The dynamic control of transmission power in VANET serves also for the optimization of the resources where it needs, can be decreased and increased depending on the circumstances of the network.

The author in this paper have found some valuable relations among different parameters of the vehicular ad hoc network. Density of the network has inverse relationship with distance on the defined highway area and distance among the moving nodes has also inverse relationship with delay constraint. It is concluded that lower transmission with lower density, there are less number of collisions to access the channel to transmit the messages. Transmission range is inversely proportional to the number of hops in the multi hop broadcast mode of the vehicular node. Broadcast storm problem decreases the reliability of the network but by using clusters, broadcast storm problem can be controlled. These parameters and their relations among themselves serve to formalize or derive the new aspects of the vehicular network particularly about the transmission power control.

**Acknowledgments**

The authors would like to acknowledge the generous support of Prof. Zoubir Mammeri and his colleagues to accomplish this paper.

**Biography**

**Muhammad Imran KHAN** is a PhD. student at University of Toulouse, France. He has done his research work in IRIT ( Institut de Recherche en Informatique et Telecommunication). He has completed his master research degree in University of Toulouse in 2008. In 2006 he has completed his bachelor in elec- trical engineering in University of Engineering and Technology Lahore, Pakistan. His current research work is in the field of wireless ad hoc networks particularly its application in vehicular ad hoc networks such as; inter vehicle communication, vehicle-to-vehicle communication and vehicle-to-infrastructure communication.